\documentclass[twocolumn]{autart}    % Enable this line and disable the
\usepackage{graphicx}          % Include this line if your
\usepackage{graphics}
\usepackage{epsfig}

\usepackage{amssymb}
\usepackage{enumerate}
\usepackage{color}
\usepackage{cases}
\newtheorem{assumption}{Assumption}
\newtheorem{lemma}{Lemma}
\newtheorem{definition}{Definition}
\newtheorem{theorem}{Theorem}
\newtheorem{remark}{Remark}

\begin{document}

\begin{frontmatter}
\title{Fixed-time consensus of multiple
double-integrator systems under directed topologies: A motion-planning approach} % Title, preferably not more
\thanks[footnotemark]{Corresponding author.  This work is supported by the National Science Foundation of China under Grants.}
\author [Nwpu]{Yongfang Liu \thanksref{footnotemark}}\ead{liuyongfangpku@gmail.com},
\author [Nwpu]{Yu Zhao}\ead{yuzhao5977@gmail.com},    % Add the
\author[Auts]{Wei Ren}\ead{ren@ece.ucr.edu},              % e-mail address
\author[Baiae]{Guanrong Chen}\ead{eegchen@cityu.edu.hk}  % (ead) as shown

\address [Nwpu]{Department of Traffic and Control Engineering, School of Automation, Northwestern Polytechnical University, Xi'an Shaanxi, 710129, China}  % Please supply          % full addresses
%\address [Pek]{State Key Laboratory for Turbulence and Complex Systems, Department of Mechanics and Engineering Science, College of Engineering, Peking University, Beijing 100871, P.R. China}  % Please supply          % full addresses
\address[Auts]{Department of Electrical and Computer Engineering
University of California,
Riverside, CA 92521, USA}
\address[Baiae]{Department of Electronic Engineering, City University of Hong Kong, Hong Kong SAR, China}

%\title{In Catilinam IV\thanksref{footnoteinfo}} % Title, preferably not more
%                                                % than 10 words.
%
%\thanks[footnoteinfo]{This paper was not presented at any IFAC
%meeting. Corresponding author M.~T.~Cicero. Tel. +XXXIX-VI-mmmxxi.
%Fax +XXXIX-VI-mmmxxv.}
%
%\author[Paestum]{Yu Zhao}\ead{yuzhao5977@gmail.com},    % Add the
%\author[Rome]{Julius Caesar}\ead{julius@caesar.ir},               % e-mail address
%\author[Baiae]{Publius Maro Vergilius}\ead{vergilius@culture.ir}  % (ead) as shown
%
%\address[Paestum]{Buckingham Palace, Paestum}  % Please supply
%\address[Rome]{Senate House, Rome}             % full addresses
%\address[Baiae]{The White House, Baiae}        % here.

                             % keyword list or with the
                                          % help of the Automatica
                                          % keyword wizard

\begin{abstract}                          % Abstract of not more than 200 words.
This paper investigates the fixed-time consensus problem under directed topologies. By using a motion-planning approach, a class of distributed fixed-time algorithms are developed for a multi-agent system with double-integrator dynamics. In the context of the fixed-time consensus, we focus on both directed fixed and switching topologies. Under the directed fixed topology, a novel class of distributed algorithms are designed, which guarantee the consensus of the multi-agent system with a fixed settling time if the topology has a directed spanning tree. Under the directed periodically switching topologies,  the fixed-time consensus is solved via the proposed algorithms if the topologies jointly have a directed spanning tree. In particular, the fixed settling time can be off-line pre-assigned according to task requirements. Compared with the existing results, to our best knowledge, it is the first time to solve the fixed-time consensus problem for double-integrator systems under directed topologies. Finally, a numerical example is given to illustrate the effectiveness of the analytical results.
\end{abstract}

\end{frontmatter}

\section{Introduction}
Over the past two decades, with the advent of wireless networks and powerful embedded systems, the distributed coordination of multi-agent systems has received significant attention in the control community due to its wide  applications in various engineering systems such as data fusion of sensor networks, task cooperation of robots, synchronization of distributed oscillators, and formation maneuver of unmanned vehicles. As the most fundamental research topic for multi-agent coordination, consensus problems have been investigated intensively. Consensus refers to a group of agents reaching an agreement on certain quantities of interest via local interaction. By specifying desired separations among different agents, consensus algorithms can be applied to achieve distributed coordination including formation control and flocking.

The consensus problems have been primarily studied for multi-agent systems with different dynamics (see \cite{Jadbabaie03,Olfati04,Ren05,Li:13,Cao08,Hong08,Hong08ifac,Fu:14,Liu152,Zhao151,Zhang:13,Liu15,Zhao13} and references therein). According to the rate of convergence, which is a significant performance index for evaluating the effectiveness
of the designed consensus algorithms, existing consensus studies can be roughly categorized into two classes, namely, asymptotic consensus and finite-time consensus. Asymptotic consensus problems were widely investigated under different scenarios \cite{Ren05,Li:13,Cao08,Hong08,Fu:14,Song10,Su11,Hu12}. In \cite{Ren05}, under directed switching topologies, asymptotic consensus problems were solved if and only if
the time-varying network topologies jointly had a directed
spanning tree. Recently, some conditions for second-order consensus were
derived in \cite{Hong08,Song10,Hu12}. By using adaptive control approaches, the adaptive consensus problem was studied in \cite{Li:13,Su11}. Furthermore, the consensus
tracking problem of multiple Euler-Lagrange dynamics was
studied in \cite{Zhao151,Meng:13}.

Different from the asymptotic consensus, achieving consensus in finite time was also studied by many researchers. The finite-time consensus problem was studied in \cite{Cortes06} for multiple single-integrator systems, where the signed gradient flows of a differential function and discontinuous algorithms were used. Since then, a variety of finite-time consensus algorithms were proposed to solve the finite-time consensus problem under different scenarios (see  \cite{Liu15,Hui09,Wang08,Li11,Chen11,Hendrickx13,Zhao16,Xu:13} and references therein). In \cite{Chen11,Hendrickx13}, the finite-time average consensus problem was investigated for multiple single-integrator systems. Further, a class of finite-time consensus algorithms for multiple double-integrator  systems were given in \cite{Hong08ifac,Zhang:13,Zhao13,Meng:13,Wang08,Li11}.  Then, the finite-time consensus problem for multiple non-identical second-order nonlinear systems was studied in \cite{Zhao16} with the settling time estimation. However, the settling time functions in \cite{Zhao16} depended on initial states of the agents, which prohibited
their practical applications if the knowledge of initial conditions was
unavailable in advance.

Recently, the authors in \cite{Zuo12} presented a
novel class of nonlinear consensus algorithms under an undirected topology for single-integrator
multi-agent networks, called fixed-time consensus which assumed
uniform boundedness of a settling time regardless of the initial
conditions.  The results in \cite{Zuo12} were further generalized in \cite{Zuoijss}
to solve the robust fixed-time consensus problems under undirected topologies for single-integrator systems with bounded input disturbances. Due to the nonlinear nature of the fixed-time algorithms, it was very difficult to generalize the existing results for first-order systems \cite{Zuo12,Zuoijss} to multi-agent systems with more complex agent dynamics. A first attempt was made in  \cite{Zuo15} for double-integrator systems. Further, in \cite{Fuscl}, a truly distributed algorithm was given under undirected topology, which depended only on the relative measurements of the neighboring agents.
Also, for multiple linear systems, the fixed-time formation problems were studied in \cite{Liu15} under an undirected complete graph.
It is worth noting that most of the above-mentioned works were derived for multi-agent
systems under undirected topologies. For the case of directed topologies, the existing algorithms in \cite{Zuo15} depended directly on the inputs of each agent's neighbors, which led to
a loop problem when there exists cycles in the graph. In practical applications, it is significant and challenging to design truly distributed
fixed-time consensus algorithm based only on the relative measurements of the neighboring agents for double-integrator
multi-agent systems under
directed topologies.
%Due to the limited bandwidth of network, it is not practical to ensure the continuity of information transmission among the neighboring agents. Many researchers have focused on consensus problems of multi-agent systems with sampled data \cite{Xie09,Gao11,Cao10,Yu11,Ma13,Huang16,Wen13}. By using zero-order hold circuit, the work \cite{Xie09} studied the consensus problems of continuous time first-order multi-agent systems via sampled control. In \cite{Cao10,Yu11}, sampled-data consensus algorithms for second-order systems  were proposed under directed and fixed topology. Further, consensus problem of second-order multi-agent systems is investigated by only using the sampled position information
%in \cite{Ma13}. Further, in \cite{Huang16}, by using past sampled position
%data, a novel consensus protocol designed for second-order linear multi-agent systems with a directed communication topology. It is worth noting that
%most of sampled-data based protocols in the above-mentioned works \cite{Xie09,Gao11,Cao10,Yu11,Ma13,Huang16} are designed under some conditions, which depends on the  lengths of sampling periods, eigenvalues of Laplacian matrix associated to communication topologies, and control gains of the protocols. It means that the protocols are restricted by some conditions coupled with sampling periods, communication topologies and controller gains.
%Generally, this
%is too strict in practical applications. Therefore, designing sampled-data based protocols for multi-agent systems under decoupling conditions is very meaningful and desirable.

Motivated by the above observations, by using a motion-planning approach, this paper investigates the fixed-time
consensus problem of double-integrator systems under directed fixed and switching topologies, respectively. The main results of this paper extend the existing works in
three aspects.
Firstly, by using a motion-planning approach, a novel framework is introduced
to solve the fixed-time consensus problems. In this framework, for double-integrator systems considered in this paper, compared with \cite{Zuo12,Zuo15,Zuoijss,Fuscl,Liu15}, a class of distributed algorithms are designed under a directed interaction topology, which has a directed spanning tree. Secondly, compared with the existing results in \cite{Zhao16}, where the finite settling time can only be estimated and related to initial conditions, in this paper, with the proposed fixed-time consensus algorithms, the settling time can be off-line pre-assigned according to task requirements. Unlike the results in \cite{Zuo12,Zuo15,Zuoijss,Fuscl}, the bounded settling time can be off-line designed in advance without estimations.
Thirdly, the algorithms designed in this paper are
based only on sampling measurements of the relative states  among its neighbors, which greatly reduces the cost of the network interaction.
To the best of authors' knowledge, it is the first time to solve the fixed-time consensus problems under directed fixed and switching topologies for double-integrator systems.

The remainder of this paper is organized as follows. The
preliminaries are given in
Section \ref{sec2}. The main theoretical results are established in
Sections \ref{sec3} and \ref{sec4}. A numerical example
is reported in Section \ref{sec5} to illustrate the
theoretical results. Concluding remarks are finally given
in Section \ref{sec6}.

%The remainder of this paper is organized as follows. The
%preliminaries are given in
%Section \ref{sec2}. Main theoretical results are established in
%Sections \ref{sec3} and \ref{sec4}. Some numerical examples
%are also reported in Sections \ref{sec3} and \ref{sec4} to illustrate the
%theoretical results. Concluding remarks are finally given
%in Section \ref{sec5}.

\section{Preliminaries}\label{sec2}

In this section, we introduce some preliminary knowledge of graph theory and matrix theory for the following analysis.

For a multi-agent system with $N$ agents, a directed graph $\mathcal{G}=(\mathcal{V},\mathcal{E})$ is used to model the interaction among these agents, where $\mathcal{V}=\{1,2,\cdots,N\}$ is the node set and $\mathcal{E}\subset \{(v_i,v_j):v_i,v_j\in \mathcal{V}\}$ is the edge set. An edge $(v_i,v_j)$ is an ordered pair of vertices in $\mathcal{V}$, which means that agent $j$ can receive information from agent $i$. If there is an edge from $i$ to $j$ , $i$ is defined as the parent node and $j$ is defined as the child node. The neighbors of node $i$ are denoted by $\mathcal{N}_i = \{j \in \mathcal{V} | (v_j,v_i)\in \mathcal{E}\}$, and $|\mathcal{N}_i|$ is the cardinality of $\mathcal{N}_i$. A directed tree is a directed graph, where every node, except for the root, has exactly one parent. A directed spanning tree of a directed
graph is a directed tree formed by edges that connect all the
nodes of the graph. We say that a graph has
a directed spanning tree if a subset of the edges forms a directed spanning tree.
The interaction topology may be dynamically changing. Therefore let
$\overline{\mathcal{G}}=\{\mathcal{G}_1,\mathcal{G}_2,\cdots,\mathcal{G}_s\}$
denote the set of all possible directed
graphs defined for the $N$ agents. In applications, the possible interaction topologies will likely be a subset of $\overline{\mathcal{G}}$. Obviously, $\overline{\mathcal{G}}$ has finite elements.
The union of a group of directed graphs $\{\mathcal{G}_{1},\mathcal{G}_{2},\cdots,\mathcal{G}_{m}\}\subset \overline{\mathcal{G}}$ is a directed graph with nodes given by $\mathcal{V}$ and edge set given by the union of the edge sets of $\mathcal{G}_{i},\; i = 1,\cdots,m$. The adjacency matrix $A$ associated with $\mathcal{G}$ is defined such that $a_{ij}=1$ if there is an edge from $j$ to $i$, and $a_{ij}=0$ otherwise. The Laplacian matrix of the graph associated with the adjacency matrix $A$ is given as $\mathcal{L} = [l_{ij}]\in \mathbb{R}^{N\times N}$, where $l_{ii}=\sum\limits_{j =1,j\neq i}^{N} a_{ij}$ and $l_{ij} =-a_{ij}$, $i\neq j$. Given a matrix $M=[m_{ij}]\in\mathbb{R}^{N\times N}$, it is said that $M$ is nonnegative if all its elements $m_{ij}$ are nonnegative, and   $M$ is positive if all its elements $m_{ij}$ are positive. Further, if a nonnegative matrix $M\in\mathbb{R}^{N\times N}$ satisfies $M\mathbf{1}=\mathbf{1}$, where $\mathbf{1}$ represents $[1, 1,\cdots,1]^T$ with an appropriate dimension, then it is said to be stochastic \cite{matrix}.

\section{Fixed-time consensus under a directed fixed topology}\label{sec3}

In this section, the fixed-time consensus for multiple double-integrator systems is studied under a directed fixed topology.

Consider the multi-agent system with $N$ agents labeled as $1, 2, \cdots,N$. The dynamics of each agent is described by
\begin{eqnarray}\label{10}
\dot x_i(t)=v_i(t), \;\; \dot v_i(t)=u_i(t),\;\;\;i=1,2,\cdots,N,
\end{eqnarray}
where $x_i(t)\in \mathbb{R}^n$ and $v_i(t)\in \mathbb{R}^n$ are, respectively, the position and velocity of agent $i$, and $u_i(t)\in\mathbb{R}^n$ the control input.

\begin{definition}(\textbf{fixed-time consensus})
For multi-agent systems (\ref{10}), the fixed-time consensus problem is said to be solved if and only if, for an off-line pre-assigned settling time $T_s>0$, for any initial conditions, the positions and velocities of multi-agent systems (\ref{10}) satisfy
\begin{eqnarray*}
&&\lim_{t\to T_s}\|x_i(t)-x_j(t)\|=0,\\
&&\lim_{t\to T_s}\|v_i(t)-v_j(t)\|=0,\;\forall i,j\in\mathcal{V},
\end{eqnarray*}
and $x_i(t)=x_j(t), \;v_i(t)=v_j(t)$, when $t\geq T_s$.
\end{definition}

The main objective of this section is to design a class of distributed
algorithms for multi-agent systems (\ref{10}) with double-integrator dynamics such that the positions and velocities of all agents in networks reach consensus in a fixed settling time, which can be off-line pre-assigned. To achieve this objective, a motion-planning approach is used to design the following algorithm
\begin{eqnarray}\label{11}
u_i(t)&=&-\frac{6(t_{k+1}+t_k-2t)}{(t_{k+1}-t_k)^3(|\mathcal{N}_i|+1)}\sum\limits_{j\in\mathcal{N}_i}\big(x_i(t_k)-x_j(t_k)\big)\nonumber\\
   &&-\frac{2(2t_{k+1}+t_k-3t)}{(t_{k+1}-t_k)^2(|\mathcal{N}_i|+1)}\sum\limits_{j\in\mathcal{N}_i}\big(v_i(t_k)-v_j(t_k)\big),\nonumber\\
\end{eqnarray}
where $i=1,2,\cdots,N,\;t_k\leq t<t_{k+1}$. The time sequence is given by $\{t_{k}=t_{k-1}+T_{k}\}$,  where $t_0=0$, $T_k=\frac{6}{(\pi k)^2}T_s, \;k=1,2,\cdots$,
and $T_s$ is a finite settling time which can be off-line pre-assigned according to task requirements.

{\remark{It is worth mentioning that the above distributed
algorithm (\ref{11}) is designed based on a motion-planning approach. Concretely, consider the cost function $
J_{k} = \frac{1}{2}\int_{t_k }^{t_{k+1}} \sum\limits_{i =1}^{N}{u_i^T(t) R_iu_i(t)dt}
$ and the associated Hamiltonian function $
H_{k}(t)= \frac{1}{2}\sum\limits_{i =1}^{N}u_i^T(t)R_iu_i(t) + \sum\limits_{i =1}^{N}(p^T_{x_i}(t) v_i(t)+p^T_{v_i}(t) u_i(t)),
$ with terminal conditions 
\begin{eqnarray*}
x_i(t_{k+1})&=&\frac{1}{|\mathcal{N}_i|+1}\bigg[\sum\limits_{j\in\mathcal{N}_i}x_j(t_k)+x_i(t_k)\bigg]\\
&&+\frac{t_{k+1}-t_k}{|\mathcal{N}_i|+1}\bigg[\sum\limits_{j\in\mathcal{N}_i}v_j(t_k)+v_i(t_k)\bigg],\\
v_i(t_{k+1})&=&\frac{1}{|\mathcal{N}_i|+1}\bigg[\sum\limits_{j\in\mathcal{N}_i}v_j(t_k)+v_i(t_k)\bigg],\;i=1,2,\cdots,N,
\end{eqnarray*}
where $p_{x_i}(t)\in \mathbb{R}^{n}$ and $p_{v_i}(t)\in \mathbb{R}^{n}$ both represent the co-states.
Solve the above optimal planing problem in light of Pontryagin¡¯s
principle \cite{optimal}. One obtains the consensus algorithm (\ref{11}) for multi-agent systems with double-integrator dynamics (\ref{10}). 
%%From the terminal conditions
%%above, one has that the algorithm (\ref{11}) is designed in order to
%%derive the state of every agent in systems (\ref{10}) to the predictive average
%%states of all its neighbors and itself at sampling instants. From
%%an intuitional point of view, after several times such motion
%%planning, the states of networked agents in systems
%%(\ref{10}) will achieve consensus.
}}
{\assumption\label{assst}{Suppose that the topology $\mathcal{G}$ among the agents is directed and has a directed spanning tree.}}

Before moving on, the following lemmas are firstly given.
\begin{lemma} \label{lemma1}
\cite{GraphTheory}
Under assumption \ref{assst}, zero is a simple eigenvalue of $\mathcal{L}$ with
$\mathbf{1}$ as an eigenvector and all of the nonzero eigenvalues are in the open right
half plane.
\end{lemma}

\begin{lemma}\label{lemm2}
\cite{Ren05}
Let $M=[m_{ij}]\in  \mathbb{R}^{N\times N}$ be a stochastic matrix. If $M$ has an eigenvalue $\lambda = 1 $ with algebraic multiplicity equal to one, and all the other eigenvalues satisfy $|\lambda| < 1$, then $M$ is SIA, that is, $\lim_{k\to \infty} M^k\to \mathbf{1}\xi^T$ , where $\xi=(\xi_1,\xi_2,\cdots,\xi_N)^T\in \mathbb{R}^N$ satisfies $M^T\xi=\xi$ and $\mathbf{1}^T\xi=1$.
Furthermore, each element of $\xi$ is nonnegative.
\end{lemma}

Then, the following theorem provides the main result in this section.

\begin{theorem}
Suppose Assumption \ref{assst} holds. For an off-line pre-assigned settling time $T_s$, the distributed algorithm (\ref{11}) solves the fixed-time consensus problem of the multi-agent system (\ref{10}) under directed fixed topologies, i.e., $\lim_{t\to T_s}\|x_i(t)-x_j(t)\|=0,\;\lim_{t\to T_s}\|v_i(t)-v_j(t)\|=0$, and $x_i(t)=x_j(t), \;v_i(t)=v_j(t)$, when $t\geq T_s$.
Further, the final consensus values $x^*(t)$ and $v^*(t)$ are given by
\begin{eqnarray}\label{finalconsensusvalue}
x^*(t)&=&\sum_{i=1}^N\xi_ix_i(t_{0})+(t-t_{0})\sum_{i=1}^N\xi_iv_i(t_{0}),\nonumber\\
v^*(t)&=&\sum_{i=1}^N\xi_iv_i(t_{0}),
\end{eqnarray}
where $x_i(t_0)$ and $v_i(t_0)$ are the initial states of the agents.
\end{theorem}
\textbf{Proof}: Firstly, we will prove that the states $x_i(t)$ at time sequence $\{t_{k}=t_{k-1}+T_{k}\}$ can achieve consensus as $k\to \infty$.
By substituting the consensus algorithm (\ref{11}) into multi-agent systems (\ref{10}), the closed-loop system can be obtained as follows:
 \begin{eqnarray}\label{10a}
\dot x_i(t)&=&v_i(t),\nonumber\\
 \dot v_i(t)&=&-\frac{6(t_{k+1}+t_k-2t)}{(t_{k+1}-t_k)^3(|\mathcal{N}_i|+1)}\sum\limits_{j\in\mathcal{N}_i}\big(x_i(t_k)-x_j(t_k)\big)\nonumber\\
   &&-\frac{2(2t_{k+1}+t_k-3t)}{(t_{k+1}-t_k)^2(|\mathcal{N}_i|+1)}\sum\limits_{j\in\mathcal{N}_i}\big(v_i(t_k)-v_j(t_k)\big),\nonumber\\
\end{eqnarray}
where $i=1,2,\cdots,N,$ $t\in [t_{k},t_{k+1}),\; k=0,1,\cdots$. Then, by integrating (\ref{10a}) from $t_k$ to $t_{k+1}$, one has
 \begin{eqnarray}\label{17a}
   x_i(t_{k+1})&=&\frac{1}{|\mathcal{N}_i|+1}\bigg[\sum\limits_{j\in\mathcal{N}_i}x_j(t_k)+x_i(t_k)\bigg]\nonumber\\
   &&+\frac{t_{k+1}-t_k}{|\mathcal{N}_i|+1}\bigg[\sum\limits_{j\in\mathcal{N}_i}v_j(t_k)+v_i(t_k)\bigg],  \nonumber\\
v_i(t_{k+1})&=&\frac{1}{|\mathcal{N}_i|+1}\bigg[\sum\limits_{j\in\mathcal{N}_i}v_j(t_k)+v_i(t_k)\bigg],\nonumber\\
&&i=1,2,\cdots,N.
\end{eqnarray}
Let $X(t_k)=(x^T_1(t_k),x^T_2(t_k),\cdots,x^T_N(t_k))^T$ and $V(t_k)=(v^T_1(t_k),v^T_2(t_k),\cdots,v^T_N(t_k))^T$.
One has
\begin{eqnarray*}
   X(t_{k+1})&=&(I_N-(\mathcal{N}+I_N)^{-1}\mathcal{L})\otimes I_n\cdot X(t_k)\\
   &&+(t_{k+1}-t_{k})(I_N-(\mathcal{N}+I_N)^{-1}\mathcal{L})\otimes I_n\cdot V(t_k),\\
   V(t_{k+1})&=&(I_N-(\mathcal{N}+I_N)^{-1}\mathcal{L})\otimes I_n\cdot V(t_k),
\end{eqnarray*}
where $\mathcal{N}=\mathrm{diag}(\mathcal{N}_1,\mathcal{N}_2,\cdots,\mathcal{N}_N)$.
Then, in the matrix form, one has
\begin{eqnarray*}
\left(
   \begin{array}{c}
     X(t_{k+1}) \\
     V(t_{k+1}) \\
   \end{array}
 \right)=
H{\otimes} I_n
\left(
   \begin{array}{c}
     X(t_{k}) \\
     V(t_{k}) \\
   \end{array}
 \right)=H^{k+1}{\otimes} I_n
\left(
   \begin{array}{c}
     X(t_{0}) \\
     V(t_{0}) \\
   \end{array}
 \right)
.
\end{eqnarray*}
where $$H=\left(
  \begin{array}{cc}
    (I_N{-}(\mathcal{N}{+}I_N)^{-1}\mathcal{L})  & (t_{k+1}{-}t_{k})(I_N{-}(\mathcal{N}{+}I_N)^{-1}\mathcal{L})  \\
    0 & (I_N{-}(\mathcal{N}{+}I_N)^{-1}\mathcal{L}) \\
  \end{array}
\right),$$
and $$H^{k}{=}\left(
  \begin{array}{cc}
    (I_N{-}(\mathcal{N}{+}I_N)^{-1}\mathcal{L})^{k}  & (t_{k}{-}t_{0})(I_N{-}(\mathcal{N}{+}I_N)^{-1}\mathcal{L})^{k} \\
    0 & (I_N{-}(\mathcal{N}{+}I_N)^{-1}\mathcal{L})^{k} \\
  \end{array}
\right).$$

Under Assumption \ref{assst}, the directed fixed topology $\mathcal{G}$ has a spanning tree. Thus, $I_N-(\mathcal{N}+I_N)^{-1}\mathcal{L}$ is a stochastic matrix. According to Lemma \ref{lemma1}, one gets that $I_N-(\mathcal{N}+I_N)^{-1}\mathcal{L}$ has an eigenvalue $\lambda_1 = 1 $ with algebraic multiplicity equal to one, and all the other eigenvalues satisfy $|\lambda_i| < 1,\;i=2,\cdots,N$. Thus, it is followed from Lemma \ref{lemm2} that for matrix $I_N-(\mathcal{N}+I_N)^{-1}\mathcal{L}$, there exists a column vector $\xi$ such that
\begin{eqnarray}\label{a}
\lim_{k\to\infty}(I_N-(\mathcal{N}+I_N)^{-1}\mathcal{L})^{k}=\mathbf{1}\mathbf{\xi}^T.
\end{eqnarray}
Besides, according to  $\{t_{k}=t_{k-1}+T_{k}\}$ and $T_k=\frac{6}{(\pi k)^2}T_s, \;k=1,2,\cdots$, one has $\lim_{k\to\infty}t_{k}= T_s$. Thus, $t_{k}{-}t_{0}$ is bounded. It follows that
\begin{eqnarray}\label{b}
\lim_{k{\to} \infty}(t_{k}{-}t_{0})[(I_N{-}(\mathcal{N}{+}I_N)^{-1}\mathcal{L})^{k}{-}\mathbf{1}\mathbf{\xi}^T]=0.
\end{eqnarray}
Denote $X^*(t )=\mathbf{1}\otimes x^*(t)$ and $V^*(t )=\mathbf{1}\otimes v^*(t)$. From (\ref{finalconsensusvalue}), one has
\begin{eqnarray*}
\left(
   \begin{array}{c}
     X^*(t ) \\
     V^*(t ) \\
   \end{array}
 \right)
&=&\Bigg[\left(
     \begin{array}{cc}
       \mathbf{1}\mathbf{\xi}^T & (t -t_0)\mathbf{1}\mathbf{\xi}^T \\
       0 & \mathbf{1}\mathbf{\xi}^T \\
     \end{array}
   \right) \otimes I_n\Bigg] \left(
   \begin{array}{c}
     X(t_{0}) \\
     V(t_{0}) \\
   \end{array}
 \right).
\end{eqnarray*}
It follows that
\begin{eqnarray*}
&&\lim_{k\to\infty}\left[\left(
   \begin{array}{c}
     X(t_{k+1}) \\
     V(t_{k+1}) \\
   \end{array}
 \right)-\left(
   \begin{array}{c}
     X^*(t_{k+1}) \\
     V^*(t_{k+1}) \\
   \end{array}
 \right)\right]\\
&=&\lim_{k\to\infty}\Bigg\{\Bigg[H^{k+1}-\left(
     \begin{array}{cc}
       \mathbf{1}\mathbf{\xi}^T & (t_{k+1} -t_0)\mathbf{1}\mathbf{\xi}^T \\
       0 & \mathbf{1}\mathbf{\xi}^T \\
     \end{array}
   \right)\Bigg]
   \otimes I_n\Bigg\}\\&&
\left(
   \begin{array}{c}
     X(t_{0}) \\
     V(t_{0}) \\
   \end{array}
 \right)
.
\end{eqnarray*}
Thus, according to (\ref{a}) and (\ref{b}), one has
\begin{eqnarray*}
\lim_{k\to\infty}\left[\left(
   \begin{array}{c}
     X(t_{k+1}) \\
     V(t_{k+1}) \\
   \end{array}
 \right)-\left(
   \begin{array}{c}
     X^*(t_{k+1}) \\
     V^*(t_{k+1}) \\
   \end{array}
 \right)\right]=0.
\end{eqnarray*}
%Note that
%\begin{eqnarray*}
%X^*(t_{k+1})&=&\mathbf{1}\otimes[(\xi^T\otimes I_n) X(t_{0})\\
%&&+(t_{k+1}-t_{0})(\xi^T\otimes I_n) V(t_{0})],\\
%V^*(t_{k+1})&=&\mathbf{1}\otimes[(\xi^T\otimes I_n) V(t_{0})].
%\end{eqnarray*}
Thus, one has the discrete states $x_i(t_{k})$ will achieve consensus with an exponential rate as $k\to \infty$, i.e., $\lim_{k\to\infty}\|x_i(t_{k})-x_j(t_k)\|= 0, \lim_{k\to\infty}\|v_i(t_{k})-v_j(t_k)\|=0,i,j=1,2,\cdots,N$.\\
Secondly, for the off-line pre-assigned  settling time $T_s$, we will prove that the discrete states $x_i(t_k)$ can achieve fixed-time consensus as $t_k\to T_s$.
Since $\lim_{k\to\infty}t_{k}= T_s$, one has
\begin{eqnarray*}
&&\lim_{t_k\to T_s}\|x_i(t_{k})-x_j(t_k)\|\\
&=&\lim_{k\to \infty}\|x_i(t_{k})-x_j(t_k)\|\\
&=&0,
\end{eqnarray*}
and
\begin{eqnarray*}
&&\lim_{t_k\to T_s}\|v_i(t_{k})-v_j(t_k)\|\\
&=&\lim_{k\to \infty}\|v_i(t_{k})-v_j(t_k)\|\\
&=&0,
\end{eqnarray*}
where $i,j=1,2,\cdots,N$. Therefore, for the off-line pre-assigned settling time $T_s$, the discrete states $x_i(t_{k})$ will achieve fixed-time consensus in exponential rate as $t_k\to T_s$.\\
Finally, we will prove that the continuous states $x_i(t)$ can achieve fixed-time consensus as $t\to T_s$. By integrating equation (\ref{10a}) from $t_k$ to $t$, it is obtained that
\begin{eqnarray}\label{10aa}
v_i(t)&=&\frac{6(t_{k+1}-t)(t_k-t)}{(t_{k+1}-t_k)^3(|\mathcal{N}_i|+1)}\sum\limits_{j\in\mathcal{N}_i}\big(x_i(t_k)-x_j(t_k)\big)\nonumber\\
   && +\frac{(4t_{k+1}-t_k-3t)(t_k-t)}{(t_{k+1}-t_k)^2(|\mathcal{N}_i|+1)}\sum\limits_{j\in\mathcal{N}_i}\big(v_i(t_k)-v_j(t_k)\big)\nonumber\\
   &&+v_i(t_k),\;
t_k\leq t<t_{k+1}.
\end{eqnarray}
Thus,
\begin{eqnarray*} &&\|v_i(t)-v_j(t)\|\\
&\leq&\frac{6(t_{k+1}-t)(t-t_k)}{(t_{k+1}-t_k)^3(|\mathcal{N}_i|+1)}\bigg\|\sum\limits_{s\in\mathcal{N}_i}\big(x_i(t_k)-x_s(t_k)\big)\bigg\| \\&&+\frac{(4t_{k+1}-t_k-3t)(t-t_k)}{(t_{k+1}-t_k)^2(|\mathcal{N}_i|+1)}\bigg\|\sum\limits_{s\in\mathcal{N}_i}\big(v_i(t_k)-v_s(t_k)\big)\bigg\|\\
&&+\frac{6(t_{k+1}-t)(t-t_k)}{(t_{k+1}-t_k)^3(|\mathcal{N}_j|+1)}\bigg\|\sum\limits_{s\in\mathcal{N}_j}\big(x_j(t_k)-x_s(t_k)\big)\bigg\| \\&&+\frac{(4t_{k+1}-t_k-3t)(t-t_k)}{(t_{k+1}-t_k)^2(|\mathcal{N}_j|+1)}\bigg\|\sum\limits_{s\in\mathcal{N}_j}\big(v_j(t_k)-v_s(t_k)\big)\bigg\|\\
&&+ \parallel v_i(t_k)-v_j(t_k)\parallel\\  &\leq&\frac{6}{(t_{k+1}-t_k)(|\mathcal{N}_i|+1)}\bigg\|\sum\limits_{s\in\mathcal{N}_i}\big(x_i(t_k)-x_s(t_k)\big)\bigg\|\\
&&+\frac{4}{(|\mathcal{N}_i|+1)}\bigg\|\sum\limits_{s\in\mathcal{N}_i}\big(v_i(t_k)-v_s(t_k)\big)\bigg\|\\
&&+\frac{6}{(t_{k+1}-t_k)(|\mathcal{N}_j|+1)}\bigg\|\sum\limits_{s\in\mathcal{N}_j}\big(x_j(t_k)-x_s(t_k)\big)\bigg\|\\&&
+\frac{4}{(|\mathcal{N}_j|+1)}\bigg\|\sum\limits_{s\in\mathcal{N}_j}\big(v_j(t_k)-v_s(t_k)\big)\bigg\|\\
&&+\parallel v_i(t_k)-v_j(t_k)\parallel,
\end{eqnarray*}
where $t_k\leq t<t_{k+1}$.
Note that $\lim_{t_k\to T_s}\|x_i(t_k)-x_j(t_k)\|=0$ with an  exponential rate and $\lim_{t_k\to T_s}(t_{k}-t_{k-1})=\lim_{k\to \infty}\frac{6}{(\pi k)^2}T_s=0$ with a polynomial rate. Thus, one has
\begin{eqnarray*}
\lim_{t_k\to T_s}\frac{6}{(t_{k+1}-t_k)(|\mathcal{N}_i|+1)}\bigg\|\sum\limits_{j\in\mathcal{N}_i}\big(x_i(t_k)-x_j(t_k)\big)\bigg\|= 0.
\end{eqnarray*}
Besides, according to $\lim_{t_k\to T_s} \bigg\|\sum\limits_{j\in\mathcal{N}_i}\big(x_i(t_k)-x_j(t_k)\big)\bigg\|=0$ and $\lim_{t_k\to T_s} \bigg\|\sum\limits_{j\in\mathcal{N}_i}\big(v_i(t_k)-v_j(t_k)\big)\bigg\|=0$, one has $\lim_{t_k\to T_s}\|v_i(t)-v_j(t)\|=0$.
Further, integrating equation (\ref{10aa}) from $t_k$ to $t$, one gets
\begin{eqnarray*}
   x_i(t)&=&-\frac{(t-t_k)^2(3t_{k+1}-t_k-2t)}{(t_{k+1}-t_k)^3(|\mathcal{N}_i|+1)}\sum\limits_{j\in\mathcal{N}_i}\big(x_i(t_k)-x_j(t_k)\big)\\
   && +\frac{(t-t_k)^2(t+t_k-2t_{k+1})}{(t_{k+1}-t_k)^2(|\mathcal{N}_i|+1)}\sum\limits_{j\in\mathcal{N}_i}\big(v_i(t_k)-v_j(t_k)\big)\\
   &&+v_i(t_k)(t-t_k)+x_i(t_k).
\end{eqnarray*}
Therefore,
\begin{eqnarray*}
   &&x_i(t)- x_j(t)\\
   &=&-\frac{(t-t_k)^2(3t_{k+1}-t_k-2t)}{(t_{k+1}-t_k)^3}\\
   &&\cdot\bigg[\frac{1}{(|\mathcal{N}_i|+1)}\sum\limits_{s\in\mathcal{N}_i}\big(x_i(t_k)-x_s(t_k)\big)\\&&-\frac{1}{(|\mathcal{N}_j|+1)}\sum\limits_{s\in\mathcal{N}_j}\big(x_j(t_k)-x_s(t_k)\big)\bigg]\\
    &&+\frac{(t-t_k)^2(t+t_k-2t_{k+1})}{(t_{k+1}-t_k)^2}\\&&
    \cdot\bigg[\frac{1}{(|\mathcal{N}_i|+1)}\sum\limits_{s\in\mathcal{N}_i}\big(v_i(t_k)-v_s(t_k)\big)\\&&-\frac{1}{(|\mathcal{N}_j|+1)}\sum\limits_{s\in\mathcal{N}_j}\big(v_j(t_k)-v_s(t_k)\big)\bigg]\\
    &&+(t-t_k)\big(v_i(t_k)-v_j(t_k)\big)+\big(x_i(t_k)-x_j(t_k)\big),\\&& t_k\leq t<t_{k+1}.
\end{eqnarray*}
Thus, one obtains
\begin{eqnarray*}
   &&\parallel x_i(t)- x_j(t)\parallel\\
   &\leq&3\bigg(\frac{1}{(|\mathcal{N}_i|+1)}\bigg\|\sum\limits_{s\in\mathcal{N}_i}\big(x_i(t_k)-x_s(t_k)\big)\bigg\|
   \\&&+\frac{1}{(|\mathcal{N}_j|+1)}\bigg\|\sum\limits_{s\in\mathcal{N}_j}\big(x_j(t_k)-x_s(t_k)\big)\bigg\|\bigg)\\
    &&+2(t_{k+1}-t_k)\bigg(\frac{1}{(|\mathcal{N}_i|+1)}\bigg\|\sum\limits_{s\in\mathcal{N}_i}\big(v_i(t_k)-v_s(t_k)\big)\bigg\|\\
    &&-\frac{1}{(|\mathcal{N}_j|+1)}\bigg\|\sum\limits_{s\in\mathcal{N}_j}\big(v_j(t_k)-v_s(t_k)\big)\bigg\|\bigg)\\
    &&+(t_{k+1}-t_k)\big\|\big(v_i(t_k)-v_j(t_k)\big)\big\|\\&&+\big\|\big(x_i(t_k)-x_j(t_k)\big)\big\|,\;t_k\leq t<t_{k+1}.
\end{eqnarray*}
Note that $t_{k+1}-t_k\leq T_s<\infty$ is upper bounded, and
$$\lim_{t_k\to T_s} \bigg\|\sum\limits_{j\in\mathcal{N}_i}\big(x_i(t_k)-x_j(t_k)\big)\bigg\|=0,$$
 $$\lim_{t_k\to T_s} \bigg\|\sum\limits_{j\in\mathcal{N}_i}\big(v_i(t_k)-v_j(t_k)\big)\bigg\|=0.$$
One has $\lim_{t_k\to T_s}\|x_i(t)-x_j(t)\|=0$.
Based on the above analysis, it follows that with the algorithm (\ref{11}), the multi-agent systems of double-integrator dynamics (\ref{10}) can achieve fixed-time consensus. The proof is completed.

\begin{remark}
Under directed topologies, the finite-time consensus problem of single-integrator multi-agent systems has been solved in \cite{Caoauto,Sayyaadi,Mauro}. However, the algorithms in \cite{Caoauto,Sayyaadi,Mauro} are difficult to develop for solving the finite-time consensus problem of double-integrator multi-agent systems under directed topologies. Also,
in \cite{Zuo12,Zuoijss}, a
fixed-time consensus algorithm is developed for integrator-type
multi-agent systems under undirected topologies.
In this paper, by using a motion-planning approach,  a novel class of distributed algorithms are proposed to solve the finite-time or  fixed-time consensus problem of double-integrator  multi-agent systems under directed topologies.
\end{remark}
\begin{remark}
Compared with the existing works \cite{Li11,Hui09,Zhang:13,Zhao16} on finite-time consensus problems and \cite{Fuscl,Zuo12,Zuoijss,Zuo15} on fixed-time consensus problems, in this paper, the  settling time can be off-line pre-assigned according to task requirements, which not only realizes the consensus in the state space but also accurately controls the settling time in the time axis.
\end{remark}

\section{Fixed-time consensus under directed periodically switching topologies}\label{sec4}
In some cases, the interaction among agents exhibits periodic phenomena, which implies that the topology among agents is periodically time-varying. Thus, we will investigate the fixed-time consensus problems of double-integrator multi-agent systems under directed periodical switching topologies. Before moving on, the following assumption is given.

\begin{assumption}\label{ass1}
For a time series $\{t_k\}$ with $t_0=0$, there exists a corresponding  directed topologies set $\overline{\mathcal{G}}=\{\mathcal{G}_0,\mathcal{G}_1,\cdots,\mathcal{G}_{m-1}\}$.
The topology among agents is periodically time-varying with the period $m$, (i.e. $\mathcal{G}_{k+m}=\mathcal{G}_{k},\;k=0,1,\cdots$, and the topologies only exist at the time instant) such that
across each time interval $[t_{k},t_{k+m-1})$, the union of the directed interaction graphs at discrete times $\{t_{k}, t_{k+1}, \cdots,t_{k+m-1} \}$ has a spanning tree.
\end{assumption}

\begin{lemma} \label{lemma4}
\cite{Ren05}
If Assumption \ref{ass1} holds, then there exists a column vector $\xi$ such that
\begin{eqnarray*}
\prod_{k=0}^{\infty}\prod_{s=0}^{m-1}[I_N-(\mathcal{N}(km+s)+I_N)^{-1}\mathcal{L}(km+s)]=\mathbf{1}\xi^T.
\end{eqnarray*}
\end{lemma}

Based on Lemma \ref{lemma4}, we will analyze the control algorithm (\ref{11}) under directed periodical switching topologies satisfying Assumption \ref{ass1}. In this case, the notation $\mathcal{N}_i$ in (\ref{11})
is replaced by  $\mathcal{N}_i(k)$.
%the following algorithm is proposed
%\begin{eqnarray}\label{11a}
%u_i(t)&=&-\frac{6(t_{k+1}+t_k-2t)}{(t_{k+1}-t_k)^3(|\mathcal{N}_i(k)|+1)}\sum\limits_{j\in\mathcal{N}_i(k)}\big(x_i(t_k)-x_j(t_k)\big)\nonumber\\
%   &&-\frac{2(2t_{k+1}+t_k-3t)}{(t_{k+1}-t_k)^2(|\mathcal{N}_i(k)|+1)}\sum\limits_{j\in\mathcal{N}_i(k)}\big(v_i(t_k)-v_j(t_k)\big),\nonumber\\
%\end{eqnarray}
%where $i=1,2,\cdots,N,\;t_k\leq t<t_{k+1}$, the time sequence is given by $\{t_{k+1}=t_k+T_{k+1}\}$,  and $T_{k+1}>0, \;k=0,1,\cdots$ can be specified according to task requirements.

\begin{theorem}
Suppose Assumption \ref{ass1} holds.
For an off-line pre-assigned settling time $T_s$, the distributed algorithm (\ref{11}) solves the fixed-time consensus problem of multi-agent system (\ref{10}) under directed periodical switching topologies.
\end{theorem}
\textbf{Proof}: Note that if we prove the discrete states $x_i(t_k)$ will achieve consensus as $k\to \infty$  with an exponential rate, then it follows from Theorem 1 that the conclusion in this theorem can be obtained. Thus, we will prove that discrete states $x_i(t_k)$ will achieve consensus as $k\to \infty$  with an exponential rate.\\
Denote $\Pi_k=\prod_{s=0}^{m-1}[I_N-(\mathcal{N}(km+s)+I_N)^{-1}\mathcal{L}(km+s)].$ For the above defined directed periodical switching topologies satisfying Assumption \ref{ass1}, one has $\Pi_{0}=\Pi_1=\cdots=\Pi_k$. Therefore, according to Lemma \ref{lemma4}, one has
\begin{eqnarray*}
&&\lim_{k\to\infty}\Pi^k_{0}\\
&=&\prod_{k=0}^{\infty}\prod_{s=0}^{m-1}[I_N-(\mathcal{N}(km+s)+I_N)^{-1}\mathcal{L}(km+s)]\\
&=&\mathbf{1}\xi^T.
\end{eqnarray*}
It follows that $\Pi^k_{0}$ will convergent to $\mathbf{1}\xi^T$ with an exponential rate as $k\to\infty$.
Thus, for multi-agent systems (\ref{10}), it follows from the proof of Theorem 1 that
\begin{eqnarray*}
&&\left(
   \begin{array}{c}
     X(t_{(k+1)m}) \\
     V(t_{(k+1)m}) \\
   \end{array}
 \right)\\
 &=&
\left(
  \begin{array}{cc}
    \Pi_{k}\otimes I_n & (t_{(k+1)m}-t_{km})\Pi_{k}\otimes I_n \\0& \Pi_{k}\otimes I_n
  \end{array}
\right)
\left(
   \begin{array}{c}
     X(t_{km}) \\
     V(t_{km}) \\
   \end{array}
 \right)\\
&=&\left(
  \begin{array}{cc}
    \Pi^{k+1}_{0}\otimes I_n & (t_{(k+1)m}-t_0)\Pi^{k+1}_{0}\otimes I_n \\0& \Pi^{k+1}_{0}\otimes I_n
  \end{array}
\right)
\left(
   \begin{array}{c}
     X(t_{0}) \\
     V(t_{0}) \\
   \end{array}
 \right)
.
\end{eqnarray*}
Note that
\begin{eqnarray*}
&&\left(
   \begin{array}{c}
     X^*(t_{(k+1)m}) \\
     V^*(t_{(k+1)m}) \\
   \end{array}
 \right)\\
&=&\left(
     \begin{array}{cc}
       \mathbf{1}\mathbf{\xi}^T & (t_{(k+1)m}-t_0)\mathbf{1}\mathbf{\xi}^T \\
       0 & \mathbf{1}\mathbf{\xi}^T \\
     \end{array}
   \right) \otimes I_n\cdot \left(
   \begin{array}{c}
     X(t_{0}) \\
     V(t_{0}) \\
   \end{array}
 \right).
\end{eqnarray*}
Similar to the proof of Theorem 1, one has
\begin{eqnarray}\label{aaaaa}
\lim_{k\to\infty}\left[\left(
   \begin{array}{c}
     X(t_{(k+1)m}) \\
     V(t_{(k+1)m}) \\
   \end{array}
 \right)-\left(
   \begin{array}{c}
     X^*(t_{(k+1)m}) \\
     V^*(t_{(k+1)m}) \\
   \end{array}
 \right)\right]=0.
\end{eqnarray}
Thus, states $x_i(t_{km})$ and $v_i(t_{km}),\;i=1,\cdots,N$, can achieve consensus as $k\to\infty$, respectively.
Besides, note that
\begin{eqnarray*}
\left(
   \begin{array}{c}
     X(t_{km+1}) \\
     V(t_{km+1}) \\
   \end{array}
 \right)&=&
\left(
  \begin{array}{cc}
    1 & (t_{km+1}-t_{km}) \\0& 1
  \end{array}
\right)\\
 &&\otimes [I_N-(\mathcal{N}(km)+I_N)^{-1}\mathcal{L}(km)]\otimes I_n\\
 &&\cdot
\left(
   \begin{array}{c}
     X(t_{km}) \\
     V(t_{km}) \\
   \end{array}
 \right).
\end{eqnarray*}
Thus, one has
\begin{eqnarray*}
&&\left(
   \begin{array}{c}
     X(t_{km+1}) \\
     V(t_{km+1}) \\
   \end{array}
 \right)-\left(
   \begin{array}{c}
     X^*(t_{km+1}) \\
     V^*(t_{km+1}) \\
   \end{array}
 \right)\\
 &=&
\left(
  \begin{array}{cc}
    1 & (t_{km+1}-t_{km}) \\0& 1
  \end{array}
\right)\\
 &&\otimes [I_N-(\mathcal{N}(km)+I_N)^{-1}\mathcal{L}(km)]\otimes I_n\\
 &&\cdot
\left[\left(
   \begin{array}{c}
     X(t_{km}) \\
     V(t_{km}) \\
   \end{array}
 \right)-\left(
   \begin{array}{c}
     X^*(t_{km}) \\
     V^*(t_{km}) \\
   \end{array}
 \right)\right].
\end{eqnarray*}
Since the boundness of $t_{km+1}-t_{km}$, it follows from (\ref{aaaaa}) that
\begin{eqnarray*}
\lim_{k\to\infty}\left[\left(
   \begin{array}{c}
     X(t_{km+1}) \\
     V(t_{km+1}) \\
   \end{array}
 \right)-\left(
   \begin{array}{c}
     X^*(t_{km+1}) \\
     V^*(t_{km+1}) \\
   \end{array}
 \right)\right]=0.
\end{eqnarray*}
Similarly, one has
\begin{eqnarray*}
\lim_{k\to\infty}\left[\left(
   \begin{array}{c}
     X(t_{km+s}) \\
     V(t_{km+s}) \\
   \end{array}
 \right)-\left(
   \begin{array}{c}
     X^*(t_{km+s}) \\
     V^*(t_{km+s}) \\
   \end{array}
 \right)\right]=0,
\end{eqnarray*}
where $s=1,2,\cdots,m-1.$
Thus, one has $\lim_{k\to \infty}\|x_i(t_{k})-x_j(t_k)\|= 0, \lim_{k\to \infty}\|v_i(t_{k})-v_j(t_k)\|= 0,\;i,j=1,2,\cdots,N$, with an exponential rate.
By the derivations similar to Theorem 1, it is easy to obtain that $\lim_{t\to T_s}\|x_i(t)-x_j(t)\|=0,\;\lim_{t\to T_s}\|v_i(t)-v_j(t)\|=0,\;i,j=1,\cdots,N$.
The proof is completed.

\begin{remark}
Note that the finite-time and fixed-time consensus problems were investigated in some interesting papers \cite{Hong08ifac,Hui09,Wang08,Li11,Zhao13,Zuo15,Liu15,Zhao16,Xu:13}. To the best of the authors' knowledge, under directed topologies, it is the first time to solve fixed-time consensus problems for multi-agent systems with double-integrator dynamics. Besides, the fixed-time algorithms designed in this paper are
based only on sampling measurements of the relative states among its neighbors, which greatly reduces the cost of the network interaction \cite{Yu11,Huang16,Wen13}.
\end{remark}
\section{Simulations}\label{sec5}
In this section, an example is given to verify the theoretical results in this paper.

%{\example{Consider a multi-agent system with 6 agents described by (\ref{1}) . The initial states are $x_1(0)=1, x_2(0)=-1, x_3(0)=6, x_4(0)=4, x_5(0)=-2, x_6(0)=-6$. For the asymptotic consensus, choose the time series as $t_k=kT+t_0$, where $T=0.25$ and $t_0=0$. Assuming the communication topology switches in a random order between the three topologies shown in Fig. 1 under Assumption 2. The simulation results are shown in Fig. 2, where it shows that the states of multi-agent system (\ref{1}) under protocol (\ref{2aa}) achieve  asymptotic consensus.
%For the finite-time case, select the settle time as $T_s=16$. In Fig. 3, the states of multi-agent system (\ref{1}) under protocol (\ref{2aa}) achieve consensus in finite time. It can been seen that the settle time is $16$s.
%}}
\begin{figure}
  \center{
  \includegraphics[width=8cm]{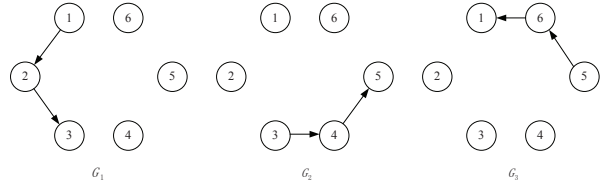}\\
  \caption{The directed periodical switching topologies among 6 agents described by (\ref{10}).} }
\end{figure}
%\begin{figure}
%  \center{
%  \includegraphics[width=8cm]{firstorder.eps}\\
%  \caption{The states of multi-agent system (\ref{1}) with motion-planning algorithm (\ref{2aa}) and $t_k=kT+t_0$, under directed and switched topology in Fig. 1.} }
%\end{figure}
%\begin{figure}
%  \center{
%  \includegraphics[width=8cm]{ffirstorder.eps}\\
%  \caption{The states of multi-agent system (\ref{1}) with  motion-planning algorithm (\ref{2aa}) and $T_k=\frac{6}{(\pi k)^2}T_s$, under directed and switched topology in Fig. 1.} }
%\end{figure}

Consider a multi-agent system with $6$ double-integrator systems described by (\ref{10}). The initial states are given by $x_1(0)=1, x_2(0)=-1, x_3(0)=2, x_4(0)=4, x_5(0)=-4, x_6(0)=-2, v_1(0)=1, v_2(0)=2, v_3(0)=3, v_4(0)=-3, v_5(0)=-2, v_6(0)=0$.
For an off-line pre-assigned fixed settling time $T_s=20$, the directed periodical switching topologies are shown in Fig. 1. The simulation results are given in Fig. 2-3, where the positions and velocities of multi-agent system (\ref{10}) under the algorithm (\ref{11}) achieve fixed-time consensus.

\begin{figure}
  \center{
  \includegraphics[width=8cm]{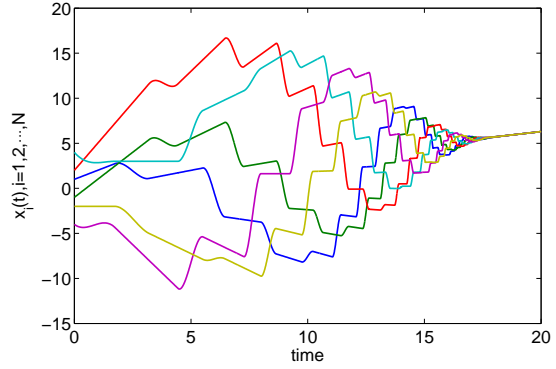}\\
  \caption{The positions of multi-agent system (\ref{10}) with  fixed-time consensus algorithm (\ref{11}) and $T_k=\frac{6}{(\pi k)^2}T_s$, under directed periodical switching topologies in Fig. 1.} }
\end{figure}
\begin{figure}
  \center{
  \includegraphics[width=8cm]{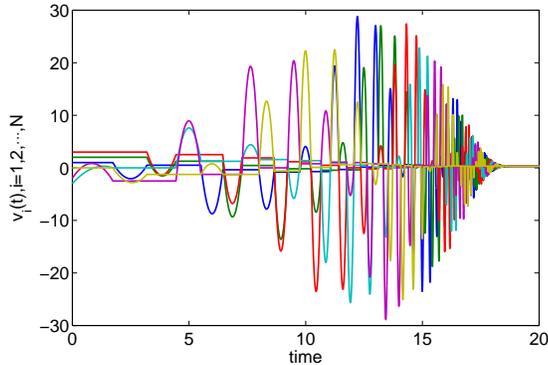}\\
  \caption{The velocities of multi-agent system (\ref{10}) with    fixed-time consensus algorithm (\ref{11}) and $T_k=\frac{6}{(\pi k)^2}T_s$, under directed periodical switching topologies in Fig. 1.} }
\end{figure}

\section{Conclusion}\label{sec6}
In this paper, the fixed-time consensus problem under directed topologies has been investigated for a group of  agents with double-integrator dynamics. By using a motion-planning approach, a class of distributed algorithms have been constructed in this paper to solve finite-time and fixed-time consensus problems under both the directed fixed and periodical switching topologies, respectively. Specially, the fixed settling time can be off-line pre-assigned. Future works will focus on solving distributed
consensus problem for multiple agents modeled by general
linear or nonlinear dynamics under directed topologies.

\end{document}